**Chapter Title:** *AI & Blockchain as sustainable teaching and learning tools to cope with the 4IR.*

*Md Aminul Islam, Data Science, University of Gloucestershire, talukder.rana.13@gmail.com*

Naahi Mumtaj Rihan, EEE, BRAC University, manjinamahi@gmail.com

***Submitted for:*** *AI + Blockchain Book, CRC Press, Under Review Now.*

**Author 1:**

Md Aminul Islam, an engineer, teacher, and researcher, studied several domains, including business, social science, education, and computer science holds BSc and MSc in Computer Science and is currently doing research in AI. Aminul has certification in education and training, networking, blockchain, and cloud and wrote 8 books for college students in Bangla. He won a few gold awards in leadership, research, and extracurricular activities. He has a membership of IEEE, the British Computer Society, the Royal Statistical Society, and STEMResearchAI. As a philanthropist through charities like Rotary International holding leadership positions from club president, DRR and MDIO Secretary is continuing to contribute to the development of the community. His focus of research is AI, ML, and Edtech.

**Author 2:**

Naahi Mumtaj Rihan is studying Electrical & Electronics Engineering at BRAC University of Bangladesh. She is passionate about creating a discrimination-free sustainable city and a world with less environmental pollution. Rihan plans to research Aerospace Engineering and work in sustainable aerospace engineering. Apart from these, she is a robotics hobbyist, sustainable, and green energy enthusiast, climate activist, thalassophile, and tech enthusiast along with having a passion for marine conversation & marine robotics.

**Abstract**

The Fourth Industrial Revolution (4IR) is transforming the way we live and work, and education. To cope with the challenges of 4IR, there is a need for innovative and sustainable teaching and learning tools. AI and blockchain technologies hold great promise in this regard, with potential benefits such as personalized learning, secure credentialing, and decentralized learning networks. This chapter presents a review of existing research on AI and blockchain in education, analyzing case studies and exploring the potential benefits and challenges of these technologies which suggests a unique model for integrating AI and blockchain into sustainable teaching and learning practices. Future research directions are discussed, including the need for more empirical studies and the exploration of ethical and social implications. By enhancing accessibility, efficacy, and security in education, AI and blockchain have the potential to revolutionize the field to ensure that students can benefit from these potentially game-changing technologies as technology develops, it will be crucial to find ways to harness its power while minimizing hazards.

**Table of Contents**

1. **Introduction:**

The serial digital mutation of industries and society comprehended as the Fourth Industrial Revolution (4IR) is fueled by the fusion of cutting-edge technologies like artificial intelligence, robots, the internet of things, blockchain, etc. It shows a reinvigorated age of economic and social change and builds on earlier industrial processes that have happened since the late 18th century.

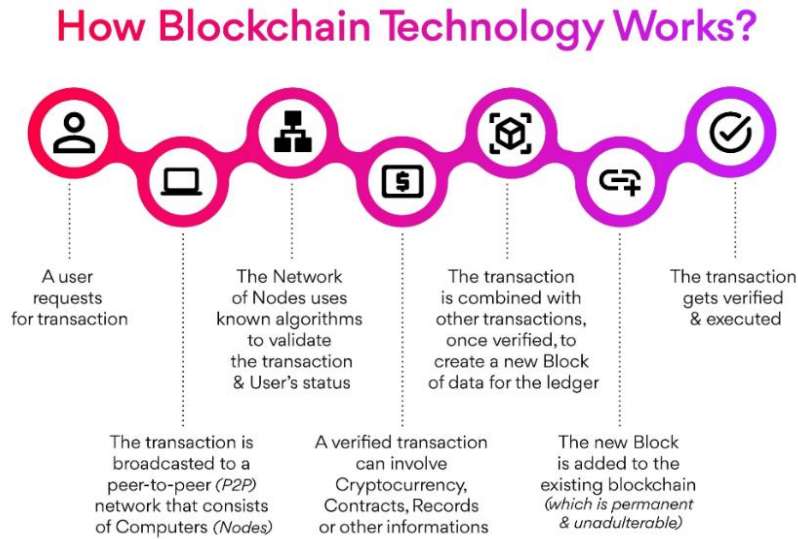

Figure01: How blockchain works (Turing, 2023)

The exponential rise of data, the outcome of progressive analytics and unctuous intelligence, the composition of new business prototypes made imaginable by digital technology, and the expanding adoption of automation and robotics in distinct industries are some of the principal drivers of the 4IR.

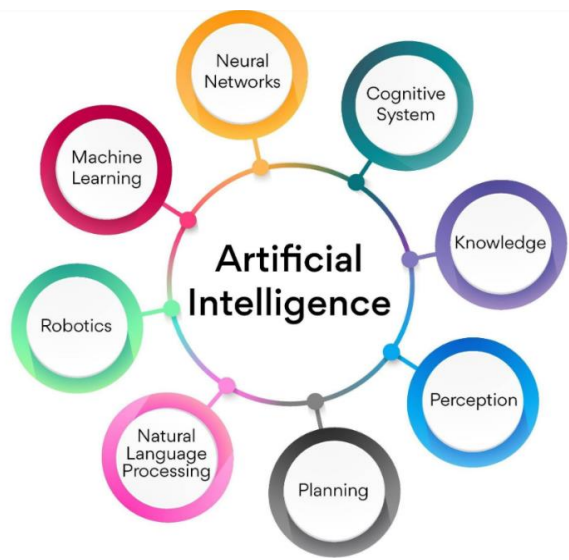

Figure 02: Different Branches of AI

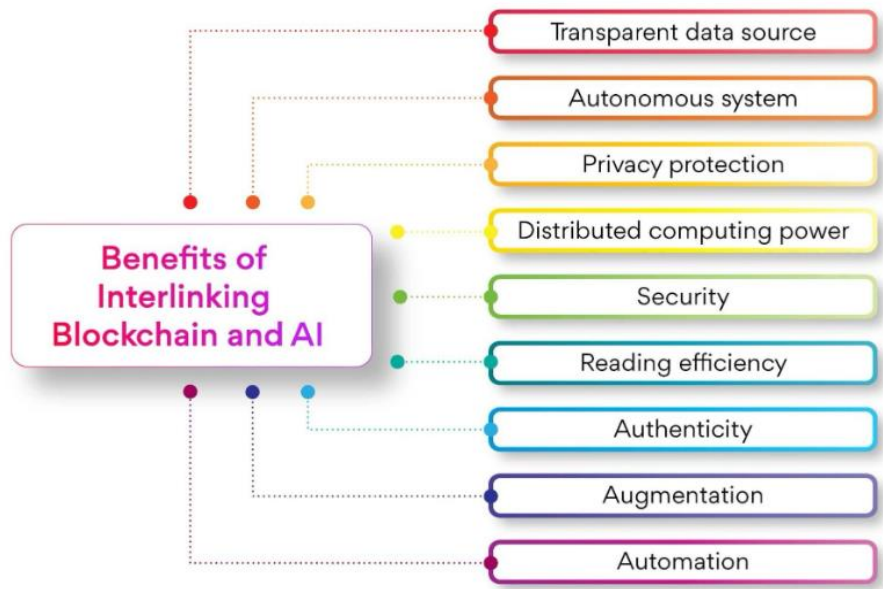

Figure 03: Interlink concept (Turing, 2023)

AI can stimulate increased production and efficiency as well as the development of renewed, cutting-edge goods and services. The use of data and AI for virtuous purposes, the prospect of growing inequality and societal division, and the leverage of automation on occupation are all significant issues that are raised by this.

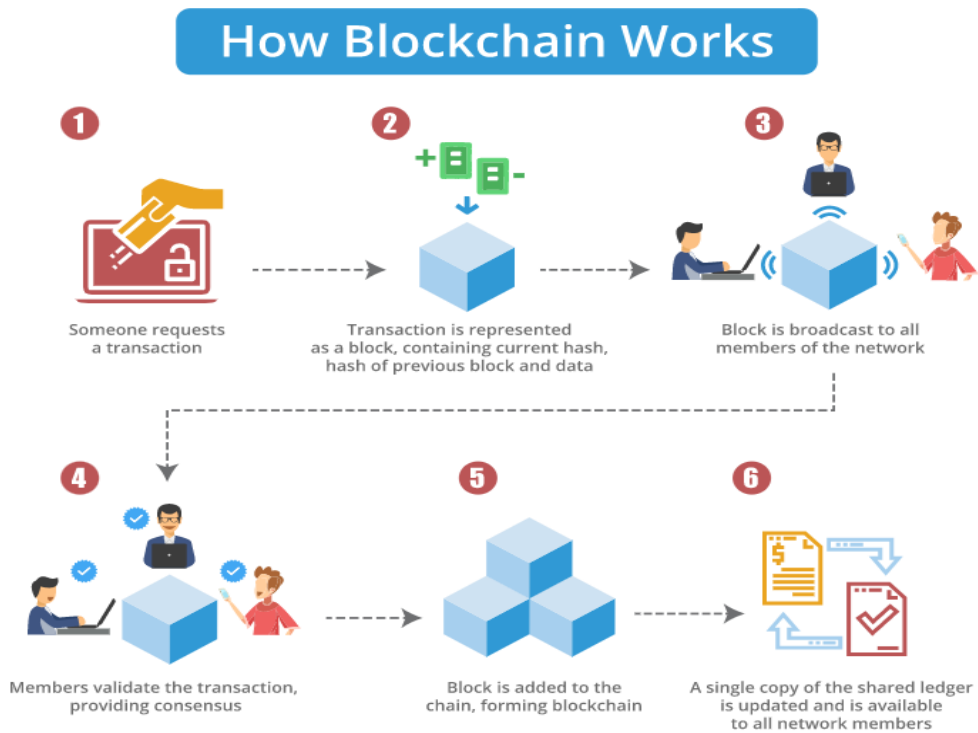

Figure 04: How blockchain works (Takyar,2023)

The 4IR signifies a substantial change in the way we view technology and its effect on the future. To certify that technology aids everyone in society, it is vital that we are acquainted with its possibilities and challenges and band as a team. It is drastically altering not only the way we work and live but also how we educate our children. The way we teach and learn is evolving as digital technologies unfold more pervasive in our daily lives, and new skills are becoming more crucial for success in the job market of the future. The boosted focus on computational reflection and digital literacy in education is one of the 4IR's major outcomes. Pupils must be proficient with digital tools and platforms, as well as have a foundational acquaintance with programming and data investigation.

New learning formats, such as online and blended understanding is another noteworthy effect of the 4IR on education. These methods supply more adaptability and accessibility, enabling learners to access understanding materials and collaborate with others wherever they may be. By operating data analytics and adaptive learning algorithms to adjust information to user needs and prerogatives, they also enable personalized learning. An interdisciplinary and project-based understanding that crystallizes on solving real-world predicaments is being placed on the curriculum because of the 4IR. This procedure promotes the blossoming of a variety of proficiencies in pupils including critical thinking, inventiveness, transmission, and teamwork—all of which are essential for success in the prospective job demand. There is a probability that numerous jobs will become ancient as automation and AI spread, and traditional educational designs may not be able to keep up with the rate of modification. This emphasizes the essence of lifelong wisdom and the elaboration of teaching to suit the directives of a world that is altering quickly (Ally and Wark, 2020).

## 2. AI and blockchain in education: An overview of the benefits and challenges:

Artificial intelligence (AI) has the potential to drastically alter education in many ways, such as streamlining administrative processes, reducing costs, and enhancing teaching and learning outcomes. The following are some potential advantages of AI in education:

- Learning Personalization: By customizing educational experiences and information to each student's specific needs and preferences, artificial intelligence can facilitate personalized learning. Additionally, it can provide adaptive feedback and tests that help students recognize their areas of strength and growth.
- Intelligent tutoring: AI-driven intelligent tutoring systems can offer students individualized feedback, direction, and support as well as help them choose which subject, they need more help in. By automating a lot of administrative responsibilities including scheduling, grading, and data analysis, teachers can save time and money while also gaining more productivity. It can identify kids who may be in danger of falling behind or dropping out and offer them specialized treatments to help them succeed using predictive analytics.

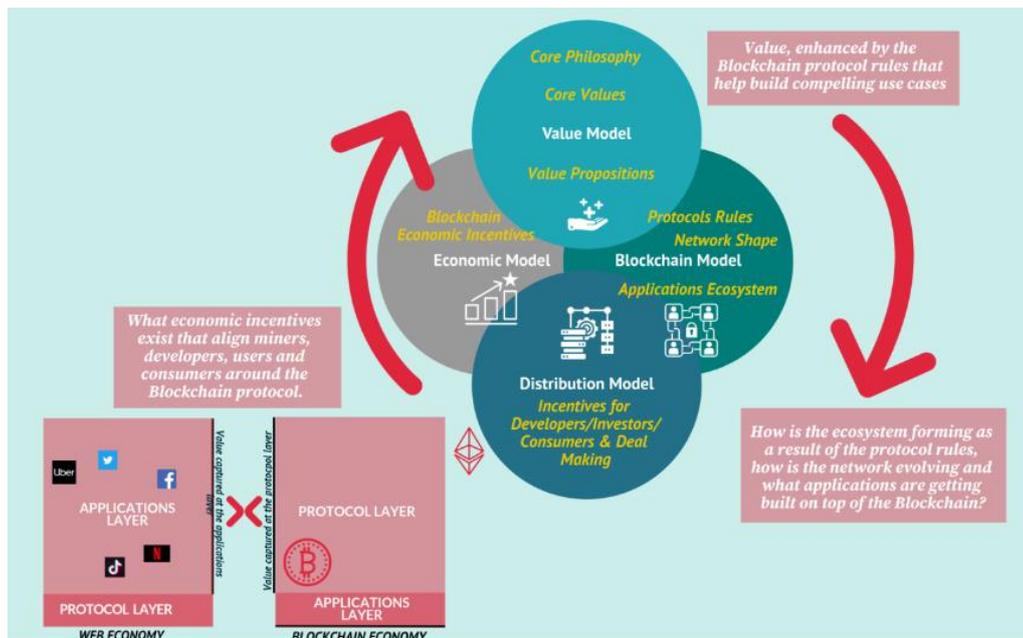
Figure 05: Blockchain in Education (Coufano, 2022)

- Digital assistants: AI capabilities can provide teachers and students with on-demand support and guidance, resolving problems and disseminating information as needed. AI can help to improve language learning by listening to speech and providing real-time feedback on pronunciation, grammar, and vocabulary using natural language processing (NLP).
- Enhancing Accessibility: Various assistive technologies like text-to-speech and speech-to-text capability, AI can improve accessibility for students with disabilities (Ouyang and Jiao, 2021).
- Improving Research: Massive amounts of data may be quickly and successfully analyzed using AI to help academics uncover patterns and insights that may not be immediately apparent using more traditional methods. (Bhaskar, 2021).

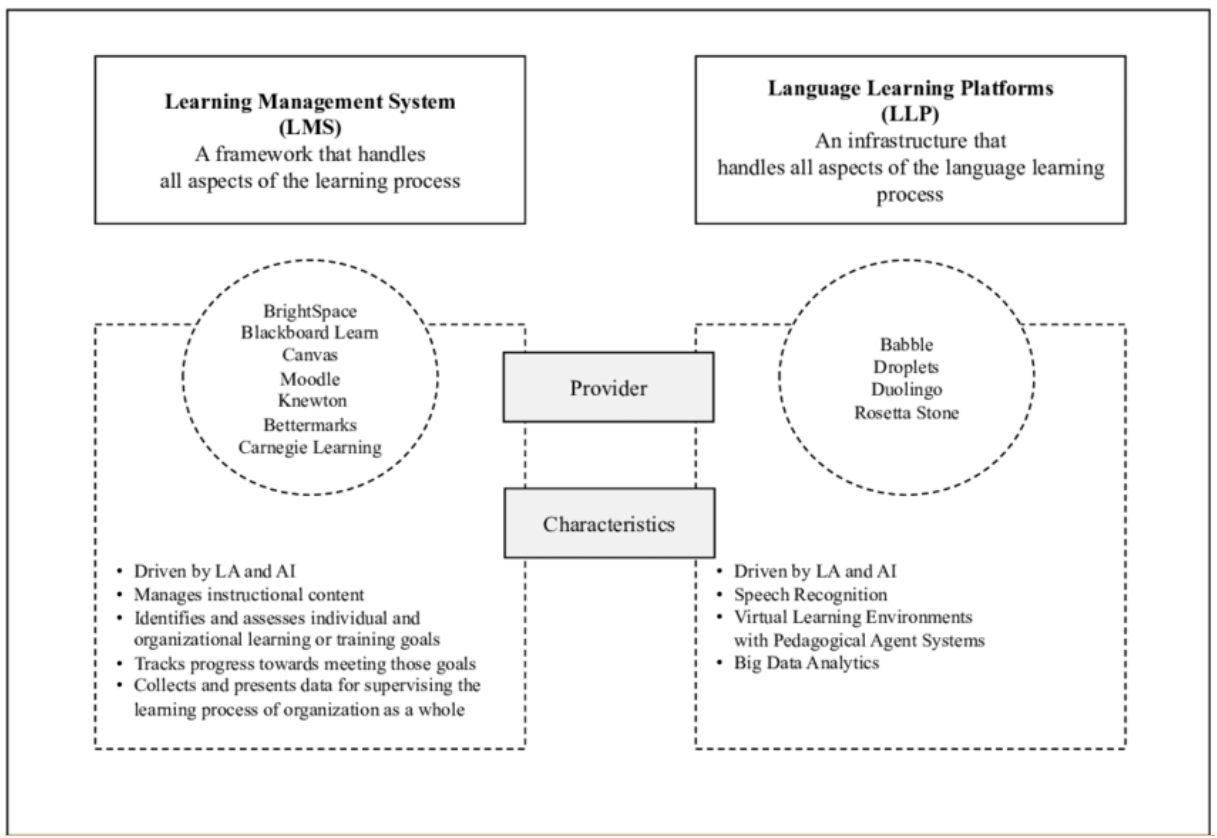

Figure 06: A new dynamic for EdTech in the age of pandemics (Hu-Berlin.de,2020)

The use of AI in education has a lot of potential advantages, but there are also hazards and difficulties that need to be considered. Racism, privacy, and the impact of automation on the workforce are a few of the issues they raise. AI has the potential to change education and how we teach and learn through careful planning and implementation.

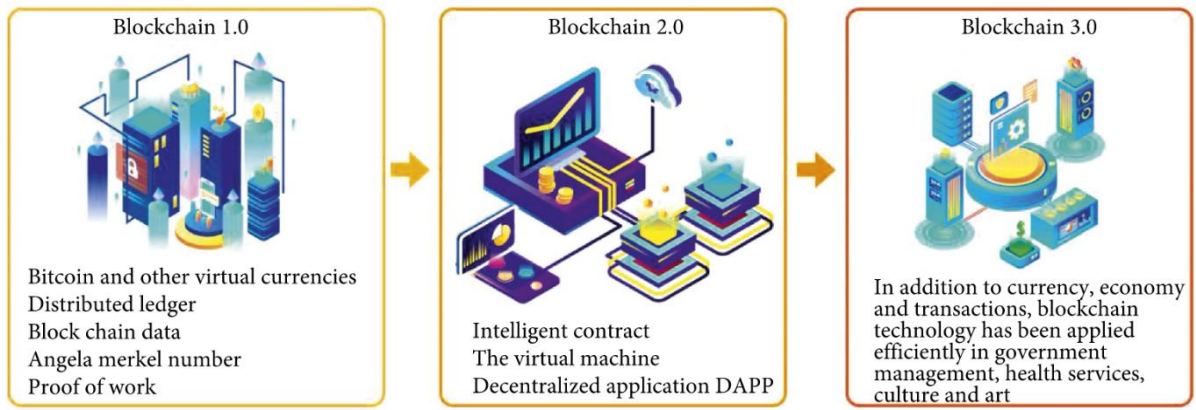
Figure 07: The development road map of blockchain technology (Chen, 2022)

The following are some potential difficulties of AI in education:

- Bias: The impartiality of AI systems can only be ensured by the data used to train them. If the data used to train AI models is biased, the AI systems that are produced can likewise be prejudiced. As AI systems have the potential to reinforce preconceptions and sustain existing disparities, this raises issues in the realm of education. Privacy and Security: Such systems could gather and keep a lot of personal data, which raises questions about security and privacy. To protect sensitive data, it is crucial to make sure AI systems are created with the proper security features.
- Cost: Adopting AI systems may be difficult for educational institutions with limited financing since they can be expensive to design and install.
- Technical Complexity: Because AI systems can be sophisticated and challenging to comprehend, it may be challenging for administrators and teachers to use them effectively.
- Ethical Concerns: Ethical concerns are raised using AI in education, including concerns about accountability, transparency, and the impact of automation on the workforce. (Pedro et al., 2019).

Now, with a safe and open system for storing and exchanging educational data and certificates, blockchain technology has the potential to revolutionize education. There are risks and issues with using blockchain in education.

Advantages:

- Secure and Immutable Records: A secure and tamper-proof solution for keeping educational documents, such as diplomas and transcripts, is provided by blockchain technology. This will help to prevent fraud and ensure that academic credentials are authentic.

- Increased Data Privacy: Blockchain technology (BCT) enables people to manage their own data and share it with others only if they have a need to know, which can help preserve student data privacy and lower the risk of data breaches.
- Decentralized Learning Networks: Building decentralized learning networks with the aid of BCT will allow students and instructors to communicate with each other directly and free of intermediaries. Learning networks might consequently improve efficiency and cost.
- Smart Contracts: Smart contracts directly encode the terms of the contract between the buyer and seller into lines of code, and these contracts carry out their own execution. They can be used in education to automate procedures like student registration, tuition payments, and the validation of academic credentials.
- Micro-Credentials: Micro-credentials can be made using BCT to acknowledge abilities or accomplishments. By sharing their credentials with potential employers and preserving them on the blockchain, people are now able to describe their skills and expertise in more depth.
- Transparent and Trustworthy Learning Ecosystems: BCT can contribute to the development of trust and accountability among students, teachers, and institutions by establishing a transparent and reliable learning ecosystem and boosting overall educational standards. (Elayyan, 2021).

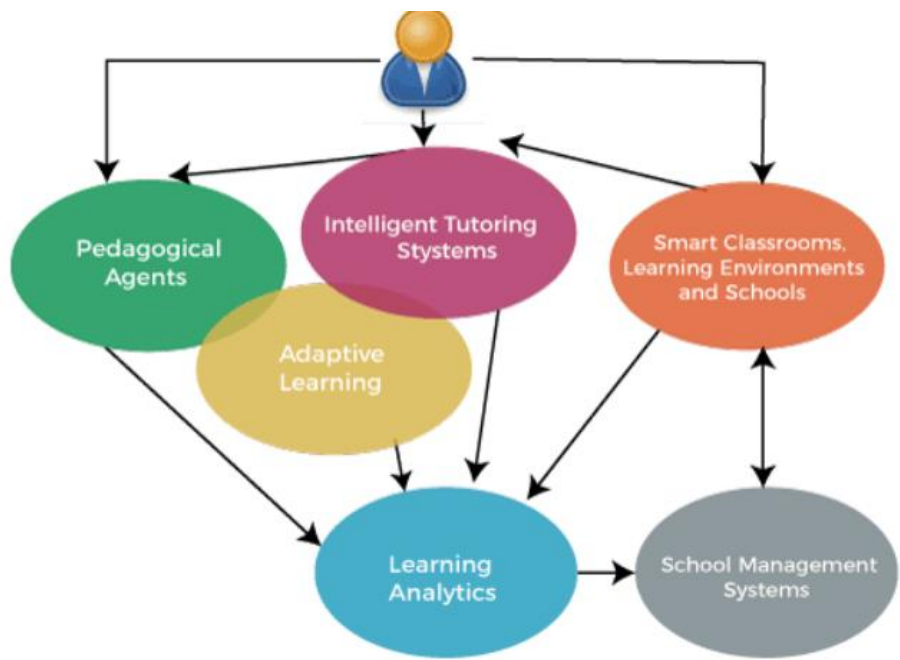

Figure 08: AI in the education system (Javapoint,2023)

Issues with blockchain in education:

- Technical Complexity: It may be difficult for administrators and teachers to properly use BCT since it might be complex and difficult to understand.
- Interoperability: The variety of existing BC platforms can make it difficult for different systems to communicate with one another. When developing a uniform system for storing and exchanging educational data, this may provide difficulties.
- Scalability: Since BCT might be resource- and time-intensive, it might be challenging to scale it to serve extensive educational networks.
- Regulatory Concerns: Using BCT in education might provide regulatory and legal challenges when it comes to issues like data protection and intellectual property.
- Cost: The significant expenses connected with its development and deployment may make adopting BCT difficult for some educational institutions. (Steiu, 2020).

### 3. AI-powered personalized learning: Customized learning experiences for learners:

The potential for individualized learning enabled by AI to totally revolutionize education is a rapidly expanding issue. To give each student a personalized learning experience, it requires using artificial intelligence algorithms to adapt instructional content to their unique requirements and preferences. AI-powered personalized learning can help students stay interested and motivated by presenting them with content that is pertinent to their interests and learning preferences, which can improve learning results.

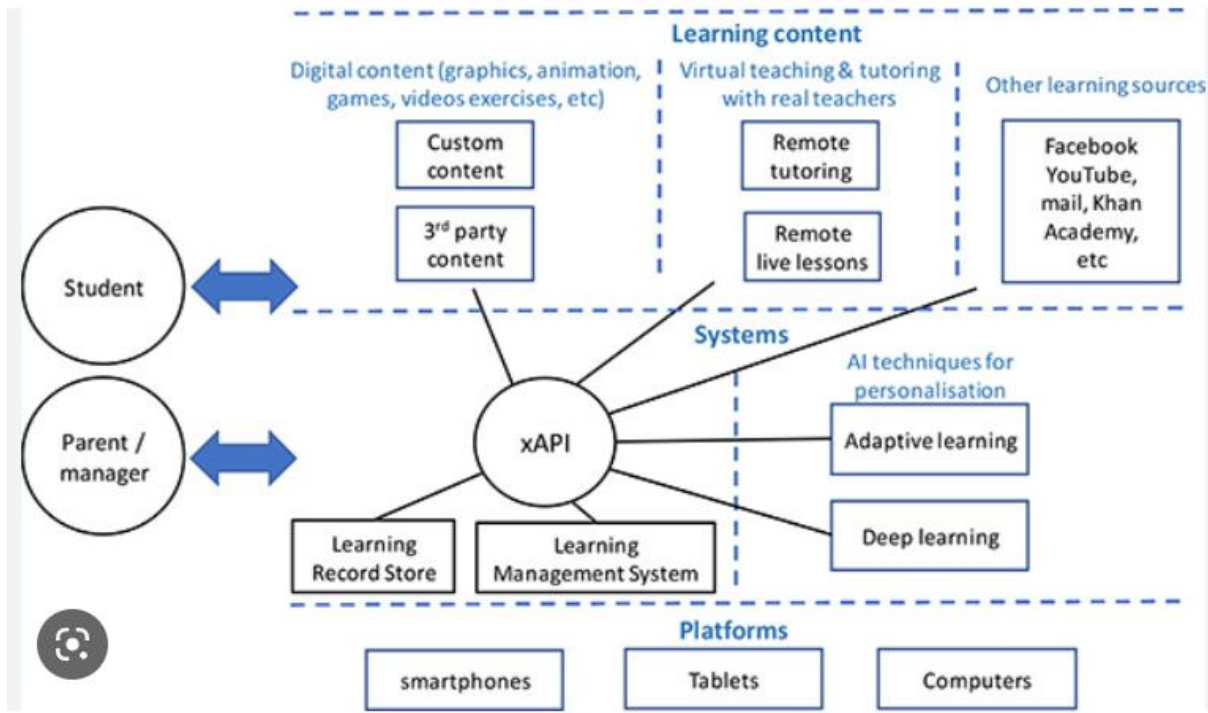

Figure 09: AI in education (Mobius Consultants, 2023)

With the use of AI-powered customized learning, the problem of student diversity may also be resolved. Teachers frequently find it difficult to satisfy the needs of students with various learning styles, experiences, and skills in typical classroom settings providing a personalized learning experience for each student that is suited to their individual needs and capabilities. AI-driven personalized learning has the potential to improve the learning process's efficacy and efficiency. AI-powered personalized learning can aid in streamlining and improving the learning process using computers to analyze student data and provide individualized suggestions. A few examples of how AI may be used to create distinctive educational experiences are shown below:

- Adaptive Learning: Adaptive learning systems driven by AI can examine student data to pinpoint areas in which they require additional assistance and suggest tailored learning resources. To ensure that each student is sufficiently challenged, these systems may also alter the content's level of difficulty in real time.
- Personalized Content: AI algorithms may help create individualized material for each student based on their learning preferences and interests. For instance, an AI-powered system may generate a student-specific reading list based on their preferred books and authors.
- Predictive Analytics: It can be used by AI to foresee a student's needs before they even materialize. AI algorithms can recognize which children are at risk of falling behind and provide specific actions to keep them on track by evaluating student data.

- Virtual Assistants: Virtual assistants powered by AI may reply to students' questions and provide them with personalized help, offering suggestions, and assisting them as they study. These assistants can modify their responses to fit the requirements of each learner.
- Gamification: Customized learning experiences that are enjoyable with gamification aided by AI. For instance, an AI-powered instructional game may modify the level of difficulty in accordance with the learner's advancement and provide specific criticism to assist them with improving.

One of the biggest challenges is the need for high-quality data. AI algorithms require access to accurate and trustworthy data to provide recommendations that are accurate. The possibility of bias, which can happen when algorithms are trained on biased data. AI-driven systems may dehumanize education by substituting automated systems for human contact (Davies et al., 2020).

### 4. Blockchain-based credentialing and certification.

A ground-breaking method for the examination and preservation of academic and professional credentials is BC-based credentialing. It entails leveraging BCT to produce impenetrable digital documents that are secure enough to share and verify.

The hassle of Verifying academic and professional qualifications has typically been a laborious and complicated process. Paper certificates and transcripts are issued by educational institutions and businesses, but these can be damaged, misplaced, or altered and need to be manually verified by employers, educational institutions, and other groups, which can be time-consuming and error prone. By creating a decentralized, impervious-to-hacking method for validating and exchanging academic and professional credentials, blockchain-based credentialing seeks to streamline this procedure.

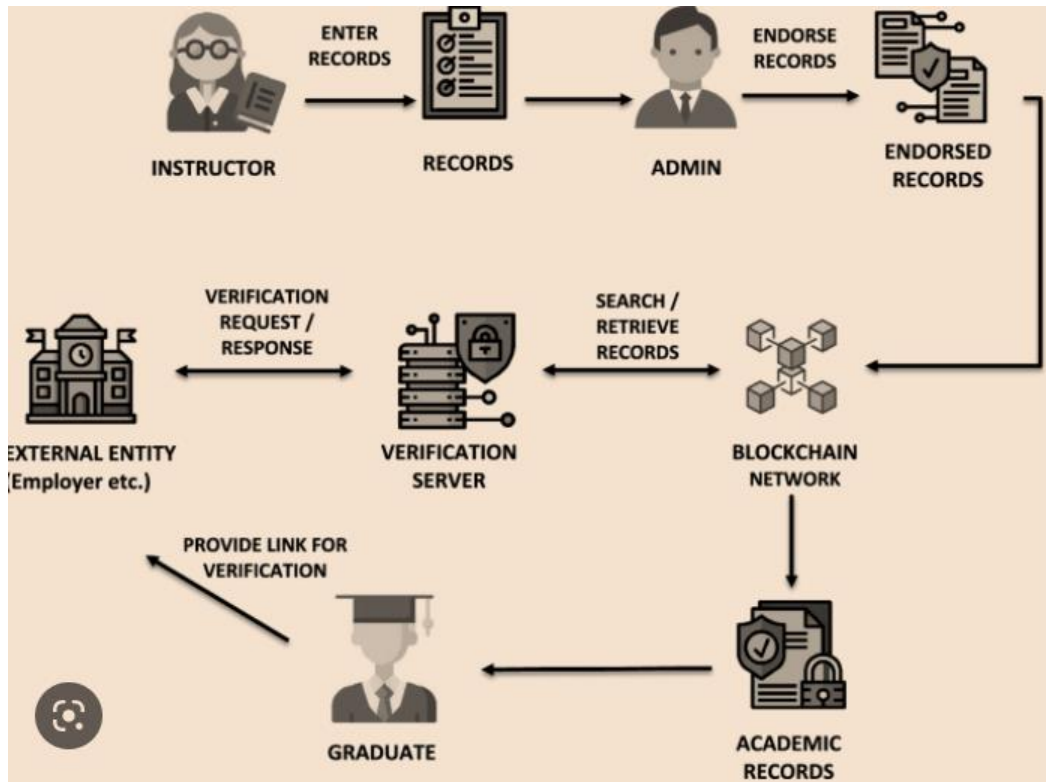

Figure 10: Blockchain-Based Academic Records Verification (Aamir et al, 2020)

The implementation of blockchain-based credentialing is not without its challenges. One challenge is the need for uniformity. For digital credentials, there must be a format that is generally recognized. if blockchain-based credentialing is to be broadly used. The requirement for privacy presents another difficulty. Despite being more secure than traditional methods, Private data exchange is still required for blockchain-based credentialing which presents privacy issues (Olaniyan et al, 2023).

Some of the main characteristics and advantages of blockchain-based certification are listed below:

- Decentralization: Blockchain-based certification is a decentralized system, which means that no single entity oversees the certification procedure. In its place, a network of nodes that verifies invalidated transactions maintains a distributed ledger.
- Immutability: Once a certificate is added to the blockchain, it cannot be modified or withdrawn. This increases the security of the certification process by removing the possibility of fraud.
- Transparency: Every transaction made on the blockchain is visible to everyone on the network and is transparent which is more transparent, which reduces the likelihood of errors and fraud.
- Efficiency: With no need for intermediaries like certification authorities, verification services, or other third-party institutions, blockchain-based certification is a more effective approach. which reduces the expenses involved with certification while also streamlining the process.
- Accessibility: The control over one's certification records is increased by utilizing blockchain-based certification. and can share them with future employers, institutions of higher learning, or other interested parties to improve their employability and mobility.

Creating secure, transparent records of diplomas and degrees using blockchains includes the following techniques:

- Creating digital records: Blockchain technology enables the production of digital records of educational credentials that are simple to access and distribute with the appropriate parties, like prospective employers or educational institutions. These documents provide information on the sort of degree or certificate received, the date of completion, and the name of the educational institution.
- Verifying records: A safe and impenetrable way for authenticating records of educational certificates is provided by blockchain technology and eliminates the potential of fraudulent activities like phony certificates or diploma mills by keeping records on a decentralized ledger.
- Providing transparency: Because all transactions on the blockchain are visible, more transparency in the certification process is guaranteed, which everyone on the network may access. As everyone can view the same data, this lowers the chance of mistakes and fraud.
- Improving efficiency: Solutions for certification based on blockchain may be more efficient since they do away with the need for middlemen and expedite the entire certification procedure. This lowers the cost of the certification procedure and gives people the need to transfer their credentials between platforms or organizations more mobility.
- Enabling self-sovereign identity: A secure, decentralized system for managing digital identities can be made using blockchain technology, giving users more control over their credentials and personal information. Such kind of self-sovereign identity can be used to confirm the identity of the certificate holder, enhancing the security and dependability of the certification process.

Using blockchain-based systems for educational qualifications is not without its difficulties, though. Systems may have trouble integrating and cooperating due to the inconsistent certification procedure, which is one of the key problems. The requirement for a trustworthy and secure digital identity system that can be used to confirm the identity of the certificate holder presents another difficulty. Some methods for using blockchain to build secure and open records of academic credentials are as below:

- Decentralization: BCT eliminates the need for a central authority to produce and manage certificates. Instead, certificates are kept on a distributed computer network, making it more challenging for fraudsters to alter the data or fabricate certificates. Decentralization enables people to manage their own certificates, providing them with greater freedom and flexibility.
- Security: A high level of security for educational credentials is provided by BCT where certificates are guaranteed to be secure and tamper-proof by using sophisticated cryptography and distributed consensus procedures. People can be assured that their certifications are safeguarded against unauthorized access or modification as a result.
- Interoperability: A universal standard for educational diplomas using BCT, many institutions and sectors will be able to distribute and recognize them easily by decreasing the complexity, this compatibility can help. and the expense of transferring certifications between various systems (Olaniyan et al,2023).

There are numerous instances of blockchain-based certification systems for education, such as the MIT Digital Diploma project, which issues and verifies digital degrees using BCT. Similarly, the Learning Machine startup has created a certification platform built on BCT that enables organizations to generate and distribute digital records of academic accomplishments.

### 5. AI-powered Assessment and Evaluation

Intelligent assessment, commonly referred to as AI-powered assessment, is a developing field that employs AI to automate and enhance the assessment process in education. Massive amounts of data may be analyzed using artificial intelligence to provide more accurate and useful analyses of student learning by utilizing algorithms of ML and NLP.

To achieve personalized learning for students, one should concentrate on educational adaptive learning technology, which uses intelligent methods to identify students' knowledge gaps and cognitive deficiencies, diagnose qualified steps for students, and thoroughly analyze data that can be designed using artificial intelligence Bayesian formulas from (1) and (2).

$$P(A|B) = \frac{P(B|A)\,P(A)}{P(B)}, \tag{1}$$

$$OP(B) = P(B|A)\,P(A) + P(B|A)\,P(A), \tag{2}$$

The difference between the learner and the behavior of the curriculum education can be created and designed; for instance, students' physiological responses like heart rate, pulse, and skin temperature can detect students' learning behavior, or students' attention distribution can be detected through mouse and keyboard input, eye movements, etc., judging students' learning and interaction data based on eye tracking such as blinking and pupil dilation.

Also. The state of the learner can be diagnosed, and future development can be predicted, which can be imported through the following formulas:

$$L = \frac{1}{2}\sum_{i=1}^{mk}\left(Y^{(K)} - T_I\right)2, \quad (3)$$

$$\left(Y^{(K)} - TI\right)^2 = \frac{1}{2}\sum_{i=1}^{MK}(\partial i)^2, \quad (4)$$

where $\partial i = Y^{(K)} - T_I$ represents the difference between the $i$ th element in any vector and the $i$ th element

$$\frac{\partial L}{\partial W} = 0, \quad (5)$$

$$W - \partial\frac{\partial l}{\partial w} = 1. \quad (6)$$

$\partial$ is the student learning rate and the step size of the weight, $(\partial l/\partial w)$ is the gradient;

The vector's i-th element; the half value is added to make the subsequent derivation calculation easier; during the design phase, the student's responses can be incorporated into the regression equation to solve the various loss functions. Moreover, the network's weights W and W can be included in the design so that the reciprocal of L and W is 0, as indicated in equations (5) and (6).

It should be noted that it can be challenging to collect student physical characteristics accurately, making it necessary in many situations to estimate them. The estimate can be compared to the likelihood and incorporated into the regression equation, for example, to analyze letter pronunciation during reading and support personalized learning. Writing proficiency may also be assessed using a tablet for children who have trouble with it, allowing them to choose more suitable learning activities. AI may identify teaching techniques and processes that are more effective for students, categorize assignments based on feedback, pay more attention to the unique needs of each student, and examine learning trends to develop students' skills (Chen, 2022)

Following are some applications of AI-driven evaluation in the classroom:

- Automated grading: One of the most common uses is automated grading for AI-powered assessment. Essays, assignments, and tests can be analyzed by AI algorithms, which can then give students immediate feedback. This can help educators save time, lighten their load, and provide more instant feedback.
- Personalized learning: With the analysis of student data and the provision of individualized feedback and suggestions, AI-powered evaluation can also be utilized to personalize learning. These tools may identify pupils' areas of weakness and provide them with tailored feedback and tools to help them develop by analyzing student performance data (Ali and Abdel-Haq, 2021).
- Adaptive testing: The creation of adaptive tests, where the test's complexity adjusts based on the student's level of knowledge, is another application of AI-powered assessment which can offer a more accurate assessment of a student's abilities and can aid in lowering test anxiety.
- Learning analytics: Assessments enabled by AI can examine large volumes of student data to provide insights into student learning. Educators might find trends and patterns in student performance data that they can use to enhance teaching and learning.
- Curriculum design: AI can be used to analyze student data, create individualized learning plans, and adapt the curriculum to the needs of each individual student. Better learning results and greater student engagement may result from this.

However, integrating AI-powered assessment in education is not without its difficulties which include biases, availability of high-quality data to train the system, etc.

**Plagiarism Detection**

The detection of plagiarism in homework assignments can be aided by AI tools. Utilizing text-matching algorithms, which compare submitted work with a sizable database of existing texts, including submissions from other students, academic journals, books, and websites, is one popular strategy. These algorithms examine the degree of similarity between the submitted work and the references already in use, identifying any possible instances of plagiarism, i.e., Turnitin. Large Language Models (LLM), NLP, etc. methods can also be used to spot irregularities in writing style, suspicious patterns, or sudden changes in vocabulary. AI systems that have been trained on a large volume of text data can identify patterns that might be signs of plagiarism, like copying and pasting text from different sources without providing proper citations or paraphrasing. Machine learning algorithms are also used by some AI-powered plagiarism detection tools to continuously increase their accuracy. By studying past instances of plagiarism, they can spot new patterns or methods used by plagiarists, adjusting, and improving their detection abilities over time.

Some components of evaluation can be automated with the use of artificial intelligence (AI), improving the effectiveness, precision, and consistency of the procedure. Using AI to automate assessment can be done in the following ways:

- Grading: AI can be used to grade multiple-choice and short-answer questions automatically which also might expedite student feedback and free up instructors' time.
- Essay evaluation: It can be used to assess essays and other writing work. The writing's structure, content, and linguistic quality can all be examined to grade written work more quickly and consistently while leaving room for arbitrary criteria like creativity and originality.
- Feedback: Students can receive quick feedback from AI on their work, enabling them to identify their errors and potential improvement areas.

The possibility of bias in the algorithms employed for automated assessment is one of the key worries and algorithms may incorporate the programmers' prejudices or may have been trained on biased data which may lead to unjust judgments and impede their ability to learn.

Another challenge is the potential for cheating by submitting work that has been generated by AI or other tools. Students may try to take advantage of the algorithms to get a good score without really doing the work themselves (Ali and Abdel-Haq, 2021). But AI-based writing detection tools can resolve this issue.

Using AI to automate evaluation can be done in the following ways:

- Performance evaluation: AI can be used to assess a person's performance at work or in school. A detailed evaluation of a person's performance may be produced by AI algorithms by accessing data from a range of sources, such as productivity indicators, attendance, and work quality. Managers and teachers can save time by doing this, and assessments will be unbiased and data-driven solutions.

- Quality control: We may evaluate the quality of products and services that can be applied in manufacturing, for instance, to find flaws in goods that are being assembled. AI can be used in customer service to assess the effectiveness of interactions between clients and service providers, for example, by assessing the tone of emails or chat messages.
- Content evaluation: The worth and application of a piece of content may be determined using AI., including blog entries, videos, and social media updates. Content producers and marketers may find this helpful as they may utilize the comments to enhance their work and more successfully reach their target audience.
- Compliance evaluation: AI can help to assess adherence to laws and standards, for instance, in the healthcare industry to assess medical records and make sure they adhere to moral and ethical norms.

**6. Blockchain-based decentralized learning networks.**

By building decentralized and secure platforms for students and instructors, BCT can completely transform the way learning networks work which makes it possible to build secure, open networks where users may communicate with one another directly and without middlemen.

- Peer-to-peer learning: Decentralized learning networks enable learners and educators to connect directly with each other, providing a peer-to-peer learning environment. So, regardless of their location or educational background, learners can gain from the knowledge of others in the network.
- Enhanced privacy: It offers improved privacy and security, safeguarding students' private data and making sure it is never shared without their permission.
- Open access: Open access to educational resources can be provided by decentralized learning networks and making it simpler for students to get the knowledge they need to succeed in their studies.
- Tamper-proof records: Using BCT, it is possible to produce tamper-proof records of academic success that anyone on the network may validate which offers a transparent and safe approach to track and validate academic credentials.

One of the biggest difficulties is the technology's intricacy, which can prevent some people from adopting it. The decentralized nature of the networks might make it difficult to maintain quality control and verify that all users are adhering to the same standards. Such technology has the potential to enable the establishment of decentralized learning networks, which could alter the way we learn and share knowledge (Alammary,2019).

The following are some ways that blockchain technology may make it possible to establish decentralized educational networks:

- Decentralized storage: With the aid of BCT, decentralized storage systems can be developed, allowing data to be kept on a network of computers rather than a single server which means that users can save and access learning materials from anywhere in the network, without relying on a centralized system.
- Immutable records: The ability to generate tamper-proof records of academic accomplishments that everyone in the network can verify is made possible by BCT and it ensures that the records are clear, correct, and cannot be changed or destroyed without authorization.
- Smart contracts: Smart contracts are carried out under the conditions of the parties' agreement and are represented by lines of code. BCT makes it possible to create smart contracts which can be used to automate some elements of education, like tracking progress, confirming task completion, and rewarding students.
- Privacy and security: By BCT, consumers' personal information is protected and ensured to never be shared without their permission. Users may share their educational resources and successes on the network.

Decentralized learning networks could have the following advantages:

- Increased accessibility: For those who do not have access to conventional educational systems, decentralized learning networks can expand educational opportunities. It generally enables instructors to reach a larger audience as well as allows students to access educational resources from anywhere in the globe.
- Cost savings: By doing away with the need for costly infrastructure like actual classrooms, textbooks, and other resources, decentralized learning networks can lower the cost of education. This may increase the cost and accessibility of education for a wider group of students.
- Improved data security: Decentralized learning networks can make sure that sensitive data, such as student's personal information and academic records, are maintained on the blockchain in a safe and tamper-proof manner in preserving the privacy of students and avoiding data breaches.

The following are some difficulties faced by decentralized learning networks:

- Technical complexity: The still developing and sophisticated BCT is the foundation for decentralized learning networks. Certain educational institutions and students may find it difficult to create and sustain a decentralized learning network.
- Adoption and scalability: Decentralized learning networks must be widely used, it can be difficult to persuade educational institutions, students, and other stakeholders to use this new technology. A network's ability to scale as its user base increases raises questions since it could have performance issues or congestion.
- Quality control: These might not have a central authority to control the caliber of exams and instructional materials. This may result in problems like erroneous or biased content or low-quality assessments, which may ultimately damage the decentralized learning network's trustworthiness.
- Legal and regulatory challenges: These systems could run into legal and administrative problems, especially when it comes to data protection and intellectual property rights. BCT may potentially bring up legal and administrative issues pertaining to data ownership and control.
- Security risks: BCT is vulnerable to hacking and other security flaws. Both educational institutions and students must take precautions to make sure that their data is shielded from modification or illegal access.

- Connectivity and infrastructure: A stable and speedy internet connection could be required for decentralized learning networks, which could not be offered everywhere. The absence of essential infrastructure, such as computers or electricity, may make it difficult for students to access decentralized learning networks.

To ensure that decentralized learning networks are available, safe, and efficient for all learners, it will be necessary for educational institutions, students, developers, and legislators to work together to overcome these obstacles (Alammary,2019).

### 7. AI-powered content creation and curation:

AI algorithms and techniques are used to create a variety of content, including text, photos, videos, and audio. In several areas, including education, this technology has the potential to transform content development. AI-powered content can quickly generate massive amounts of material which can be helpful in the educational setting, as teachers and instructional designers frequently need to produce a substantial amount of learning materials to satisfy the various demands of students.

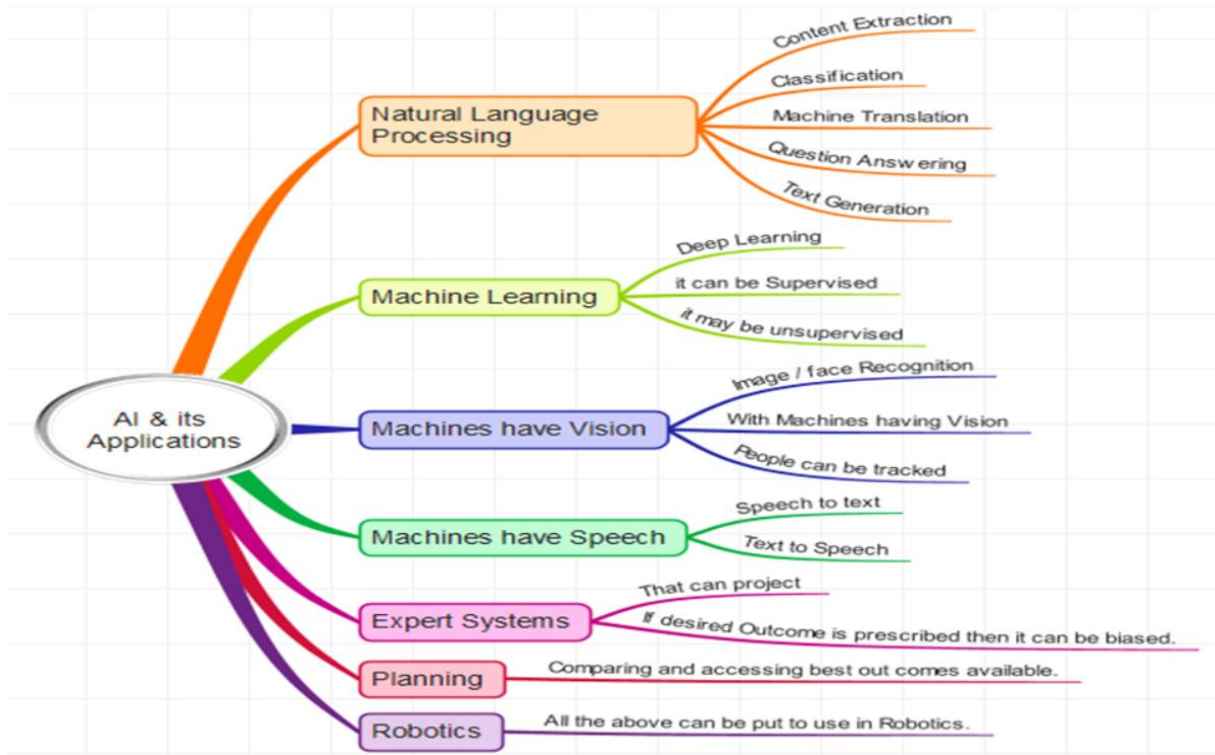

Figure 11: AI in Education landscape (Little One The Jaipuria Preschool, 2023)

The quality and accuracy of instructional content can both be increased with the use of AI-powered content development. AI can make sure that written text is clear and grammatically correct by employing NLP techniques. Data analysis and insight-providing capabilities of AI-powered content production platforms can assist teachers and instructional designers in producing more effective teaching materials. Another benefit of AI-powered content creation is its ability to customize content to the needs and preferences of certain learners. AI may create specialized learning materials that are adapted to the individual needs of each student by analyzing data on learner behavior and performance.

AI-generated content can be unoriginal and uncreative. Although AI can produce content rapidly and effectively, it might not be able to do so in a way that is truly novel or ground-breaking. This might be a problem in the classroom, where students might gain from being exposed to fresh, original ideas. AI-powered curation uses algorithms and methods to automatically sort, arrange, and suggest content based on user interests and behavior. The way learning resources are chosen and provided might be completely changed by this technology.

Personalization of the learning experience through AI-powered curating is one of its primary advantages since AI algorithms can make content recommendations that are pertinent to a student's interests and learning objectives by examining data on learner behavior and preferences. As they are more likely to find the subject engaging and valuable, this can help students stay motivated and engaged.

AI-powered curation may provide insights into learner behavior and performance and can find patterns and trends in how learners interact with curated content, which they can then use to enhance the efficacy of the next curation initiatives.

Scalability is one of the key advantages of content creation powered by AI. In poor nations or for students who might not have access to high-quality educational resources, this can increase access to and affordability of education. The caliber of educational resources can be raised using AI in content creation. For instance, chatbots powered by AI can simulate discussions with learners, assisting them in honing and practicing their language abilities. Virtual reality (VR) settings driven by AI can give students dynamic, hands-on experiences that are not achievable with conventional instructional tools.

The requirement for human oversight and involvement presents another difficulty. The text produced by the algorithm must be reviewed and edited by human professionals. Critics are concerned that the use of AI tools in education could diminish students' capacity for critical and analytical thought throughout the learning process, these tools may unintentionally discourage students from utilizing critical thinking and problem-solving skills. The need for students to conduct research and exercise critical thinking is frequently diminished by AI tools that provide ready-made answers or solutions. This reliance on AI-generated responses may inhibit their ability to analyze complex problems, independently evaluate information, and generate original solutions. To provide students with a well-rounded education that will prepare them for the challenges of the future, it is necessary to establish a balance between the development of critical thinking skills and the use of AI tools. The ability of AI-powered material curation to save time for teachers and students is another prime advantage. Learners can rely on AI algorithms to suggest top-notch materials that are pertinent to their needs rather than spending hours searching for educational resources. Teachers can utilize AI-powered curation to find resources that are compatible with their learning goals, saving time and effort. The recommendations made by the algorithm may be biased if the data used to train the algorithm was skewed in any way which may expose students to information that is erroneous or biased.AI algorithms might offer helpful recommendations, it is crucial to make sure that the suggested resources are precise, pertinent, and in line with learning goals.

Changes in instructional design, teaching strategies, and teacher preparation are just a few of the ways that the use of AI in education will have a big impact on teaching. Here are a few examples of how AI might affect education:

- Personalized learning
- Adaptive learning
- Data-driven instruction
- Automated grading
- Intelligent tutoring systems

Now the application of AI in education does, however, present certain potential difficulties. For instance:

- Concerns about bias: AI algorithms' quality is based on the data that was used to train them. The recommendations or judgments that the algorithm makes may be prejudiced if the data utilized to train the algorithm is biased in some way. It's crucial to make sure AI is applied fairly, transparently, and with as little chance of prejudice as possible.
- The need for teacher training: Teachers must possess a specific level of technological knowledge and skill to deploy AI in the classroom. To successfully integrate AI into their teaching practices, teachers may require help and training.
- The risk of dehumanization: Aspects of education that can be automated by AI include grading and assessment. And while doing so can save time and effort, It is essential for using AI in a way that complements rather than displaces human participation and teaching.

Here are some ways that learning may be affected by AI:

- Personalized Learning
- Adaptive Learning
- Improved Student Engagement
- Accessibility
- Real-time feedback
- Enhanced Collaboration
- Time Management

Yet, after all these discussed above there are some issues and difficulties with using AI in teaching also.

- Bias: Because AI algorithms have the potential to be prejudiced, different student groups may experience unequal learning outcomes.
- Dependence on Technology: Overreliance on technology among students may result in a dearth of creativity and critical thinking abilities.
- Privacy and Security: Since student data may be subject to hacking and other online dangers, the use of AI in education poses privacy and security concerns.
- Implementation: Infrastructure, support, and training must be heavily invested in for AI in education to be successful.
- Job Displacement: By using AI to automate some duties, the educational system may become unbalanced, and teachers' jobs may be lost (Ahmed and Ganapathy,2021).

## 8 Case studies: Examples of AI and blockchain in education, and their impact on teaching and learning-

Below are some studies and illustrations of the use of blockchain and AI in education:

Smart Degrees in Malta: To introduce "Smart Degrees," the Malta government worked with Learning Machine, a blockchain technology business. This approach safeguards academic credentials by storing them on a blockchain so they can't be changed or tampered with. In various universities in Malta, the system has been put in place, and students can access their credentials using a safe digital wallet.

Carnegie Mellon University's AI Tutor: An AI-powered teaching system that offers pupils individualized feedback and direction was developed by Carnegie Mellon University. The "ALEKS" system adapts to each student's learning style and speed, giving them content that is specifically designed to help them increase their knowledge and comprehension (Koedinger et al., 1997).

The University of Bahrain's Blockchain-based system: A blockchain-based system was put in place by the University of Bahrain to store and distribute student records and diplomas. Employers and other educational institutions may verify the data thanks to this method, which makes sure it can't be tampered with (JIbrel, 2019).

Also, the impact of AI on teaching and learning is seen in the following case studies:

Carnegie Learning: AI is used by Carnegie Learning to give pupils individualized learning opportunities. The technology adapts to the learning style and pace of each learner and uses machine-learning algorithms to spot places where they need more help. As a result, student outcomes have significantly improved; according to one study, students who use Carnegie Learning performed 23% better on examinations than those who used conventional techniques.

Coursera: An online learning platform called Coursera employs AI to customize each student's educational experience. Using machine learning techniques, the system evaluates each student's academic performance, then offers tailored suggestions for programs and resources. Higher engagement and completion rates have resulted from this; according to one study, students utilizing Coursera were 21% more likely to finish a course than those using conventional means.

Here are a few case studies that demonstrate how BCT affects teaching and learning:

ML: Blockchain-based digital credentials can be created and issued using software from Learning Machine. The Massachusetts Institute of Technology (MIT) is one of its customers, and it has been using ML technology to provide graduates with digital certificates since 2018. The verification process is accelerated, made safer, and made more visible using blockchain-based credentials.

Open University: Since 2015, The Open University in the United Kingdom has used blockchain to provide graduates with digital badges. The badges, which students can publish on social media and professional networking websites, signify the abilities and knowledge that they have acquired via their courses. Using blockchain-based badges has allowed The Open University to improve its standing and attract more students.

## 9 Challenges of AI and Blockchain- The demerits of these in teaching and learning:

The use of AI and blockchain in education could have certain negative effects, even though they have the potential to change teaching and learning. The following are some drawbacks of blockchain and AI in education:

- Lack of Human Interaction
- Bias and Inequality
- Cost
- Data Privacy and Security
- Complexity

It has become a common concern for education providers and teachers regarding LLM-based AI tools which can be used for cheating in assessment, homework, assignment etc. Even some creative works like writing songs, poems, articles, paintings, and drawings. It can be used for primary-level idea generation but require an adequate level of research, fine-tuning, and in-depth understanding to avoid paraphrasing and AI detection error (IOE London Blog,2023). ChatGPT can generate too generic answers to different types of questions but those are neither always correct nor high standards (Code Today, 2022).

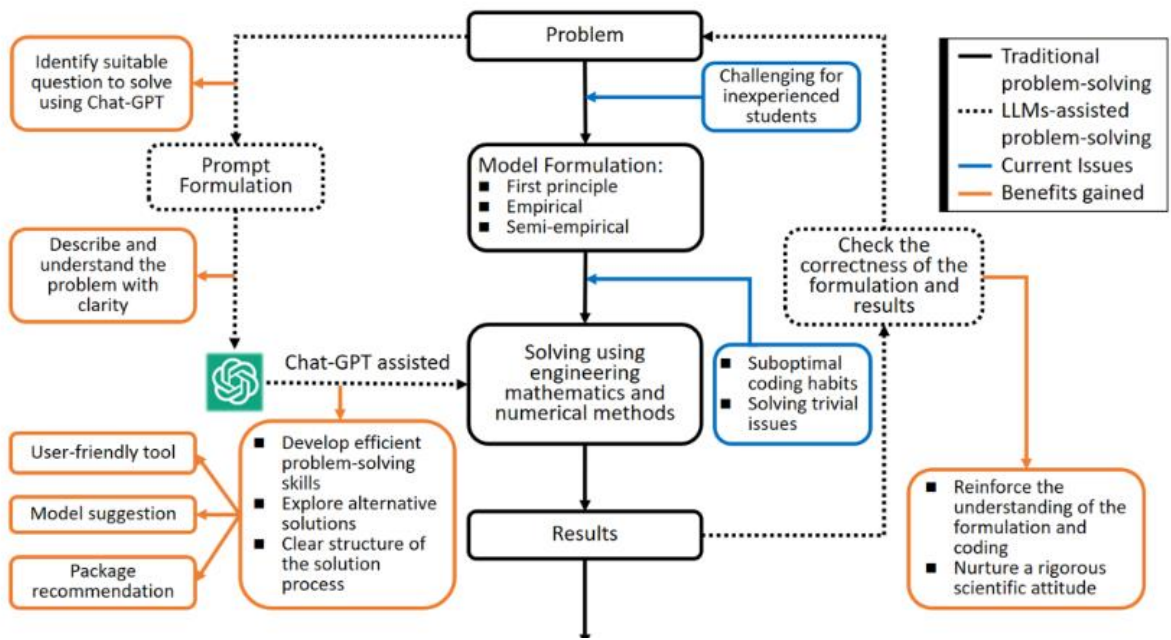

Figure 11: LLM based approach for problem solving (Tsai, M.L., et. al., 2023)

GPT and LLM models can scale up the problem-solving strategies more faster and engaging as a form of chat to formulate the ideas to narrow down.

## 10 Conclusion

Personalized, decentralized, and secure learning networks can be created by integrating blockchain and AI into teaching and learning, which has the potential to completely change the way that education is provided. While blockchain-based credentialing and certification can offer a transparent and unchangeable record of learners' accomplishments, AI-powered personalized learning can adapt educational experiences to individual needs. Whereas. blockchain can make it easier to create decentralized learning networks, Certain components of evaluation and content creation can be automated with AI. Some of the challenges that must be solved include the need for appropriate standards and regulations as well as moral concerns with data privacy. Future studies should concentrate on examining the possible advantages and disadvantages of these technologies and creating best practices for their adoption in the classroom.